\documentclass[reprint,prl,showpacs,superscriptaddress,floatfix,letterpaper]{revtex4-1}
\usepackage[colorlinks, hyperfigures, linkcolor=black, citecolor=black, breaklinks,]{hyperref}
\usepackage[centertags]{amsmath}
\usepackage{amssymb}
\usepackage{amsthm}
\usepackage[normalem]{ulem}		
\usepackage{fancyhdr}
\usepackage{graphicx,subfigure}
\usepackage{epsfig}
\usepackage{bbm}

\begin{document}

\title{Self-averaging fluctuations in the chaoticity of simple fluids}

\author{Moupriya~Das}
\affiliation{Department of Chemistry,\
  University of Massachusetts Boston,\
  Boston, MA 02125
}
\author{Jason~R.~Green}
\email[]{jason.green@umb.edu}
\affiliation{Department of Chemistry,\
  University of Massachusetts Boston,\
  Boston, MA 02125
}
\affiliation{Department of Physics,\
  University of Massachusetts Boston,\
  Boston, MA 02125
}
\affiliation{Center for Quantum and Nonequilibrium Systems,\
  University of Massachusetts Boston,\
  Boston, MA 02125
}

\begin{abstract}

Bulk properties of equilibrium liquids are a manifestation of intermolecular
forces. Here, we show how these forces imprint on dynamical fluctuations in the
Lyapunov exponents for simple fluids with and without attractive forces. While
the bulk of the spectrum is strongly self-averaging, the first Lyapunov
exponent self-averages only weakly and at a rate that depends on the length
scale of the intermolecular forces; short-range repulsive forces quantitatively
dominate longer-range attractive forces, which act as a weak perturbation that
slows the convergence to the thermodynamic limit. Regardless of intermolecular
forces, the fluctuations in the Kolmogorov-Sinai entropy rate diverge, as one
expects for an extensive quantity, and the spontaneous fluctuations of these
dynamical observables obey fluctuation-dissipation-like relationships.
Together, these results are a representation of the van der Waals picture of
fluids and another lens through which we can view the liquid state.

\end{abstract}

\maketitle

\textit{Introduction.--} Liquids live in a state between the structural order
typical of solids and the dynamical randomness of gases~\cite{barker2,inge}.
While molecular chaos underlies the existence of gaseous thermodynamic states,
this notion, and Boltzmann's \textit{Stosszahlansatz}~\cite{boltz}, are absent
from the theories of liquids~\cite{hansen}. The coarse features of simple
fluids are instead seen as a result of the forces between their molecular
constituents~\cite{widom}. Since van der Waals, the prevailing view has been
that strong repulsive forces determine the structural arrangements of molecules
in a liquid away from the critical point. Weak, longer-range attractive forces,
though, have relatively little effect~\cite{van,voll}. This paradigm forms a
basis for the statistical mechanics of equilibrium
liquids~\cite{fisk,barker,barker1,chand} and, more recently, has been
considered for the slow dynamics of supercooled liquids~\cite{berth,dyre,dell}.
However, this view neglects the fluidity of liquid matter that comes from
the incessant thermal motion of molecules, motion that, because of these
intermolecular forces, is intrinsically chaotic~\cite{green1}. It remains an
open question how the liquid state emerges from the Lyapunov instability and
deterministic chaos of molecular dynamics~\cite{bose} and whether the van der
Waals picture mirrors a dynamical perspective, where attractive and repulsive
interactions play distinct roles in the emergent chaotic behavior of
equilibrium liquids. In this Letter, we address this question.

A signature of the nonlinear dynamics of fluids, and any dynamical system
exhibiting deterministic chaos, is the divergence of initially close phase
space trajectories~\cite{gaspard}. Finite-time Lyapunov exponents (FTLEs) are
the exponential rates of divergence and important measures of the sensitivity
to initial conditions characteristic of chaos. While their fluctuations on
finite-time scales ultimately decay in the long-time limit~\cite{piko,lopez},
it is unknown how these fluctuations scale in the thermodynamic limit for
fluids. This fact is largely because FTLE calculations are limited to systems
that are small or restricted to one or two spatial dimensions~\cite{cost}. By
overcoming this challenge, we show that the finite-size scaling of these
fluctuations, that is, how they ``self-average,'' is a quantitative
representation of the van der Waals picture.

Thermodynamic states are generally a consequence of the self-averaging of
microscopic properties. A system self-averages its independent subsystems if a
global observable is an average of that observable over the independent
subsystems. More precisely, a system is self-averaging with respect to a
property $X$ if the relative variance $R_X=(\left<X_N^2\right> -
\left<X_N\right>^2)/\left<X_N\right>^2 \to 0$ as the system size $N\to\infty$,
with averages over statistical samples, independent subsystems, noise,
disorder, or independent time windows~\cite{lifs}. For equilibrium systems, the
relative variance is $\mathcal{O}(N^{-1})$ but more generally can be $R_X
\simeq N^{-\gamma}$ with a wandering exponent $0 \leq \gamma \leq 1$.

Through finite-size scaling of Lyapunov exponent fluctuations and numerical
estimates of $\gamma$, dissipative and Hamiltonian dynamical systems have begun
to collect into universality classes. Previous work on dissipative,
spatially extended dynamical systems in one and two dimensions has shown the
self-averaging behavior of the largest Lyapunov exponent is not only dependent
on the number of spatial dimensions but also distinct from the bulk of the
Lyapunov spectrum. For these dissipative systems, the Lyapunov exponents are
weakly self-averaging, $\gamma < 1$~\cite{kupt,lopez1}, with the dynamics of
the first Lyapunov vector belonging to the Kardar-Parisi-Zhang (KPZ)
universality class~\cite{kardar} and a $\gamma$ for the associated exponent
that is related to the known critical exponents. The bulk exponents, however,
belong to another, still unknown, universality class. There are one-dimensional
Hamiltonian systems, such as the FPU-$\beta$ and $\Phi^{4}$ models, that stand
in stark contrast. Fluctuations of the maximum Lyapunov exponent in these
models are not self-averaging but instead diverge in the thermodynamic
limit~\cite{lopez}. It is not yet clear whether this non-KPZ
behavior~\cite{piko1,rome} is a sole consequence of long-range spatiotemporal
correlations or a more universal feature of Hamiltonian dynamics. By extending
the finite-size scaling of the fluctuations of the Lyapunov spectrum to
three-dimensional Hamiltonian systems, we show here the divergence is not
universal, thus confirming the hypothesis of long-range correlations.

\begin{figure}
\includegraphics[width=0.99\columnwidth,angle=0,clip]{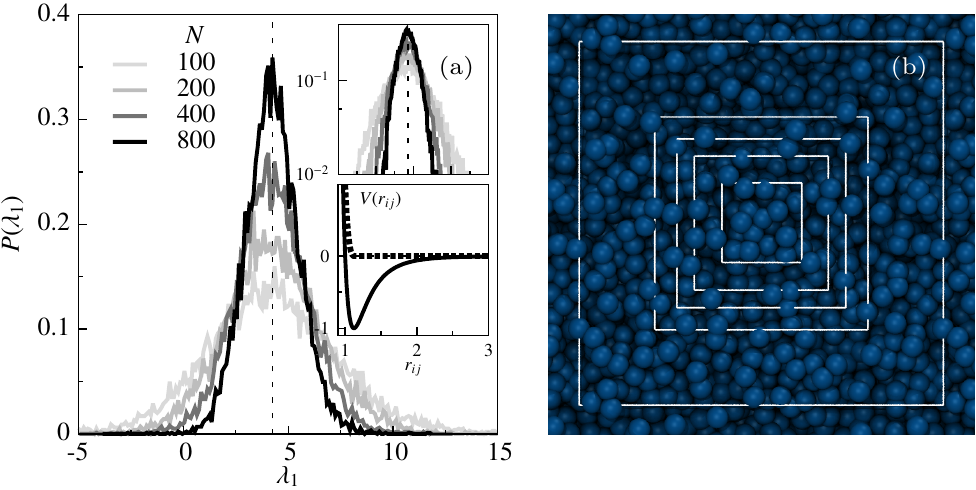}

\caption[]{\label{Fig6} (a) Empirical probability density functions of
$\lambda_{1}$ from trajectory ensembles of Weeks-Chandler-Andersen fluids with
varying numbers of particles $N=$ 100, 200, 400 and 800 at the kinetic
temperature $T=0.9$ and number density $\rho = 0.75$. One inset is a semilog
plot of the same data. The other inset shows the potential energy $V(r_{ij})$
for the interaction between Lennard-Jones and Weeks-Chandler-Andersen particles
$i$ and $j$ as a function of the interparticle distance $r_{ij}$. (b) Snapshot
showing 5 out of the 16 system sizes we simulate with $NVE$ molecular
dynamics.}

\end{figure}

\textit{Models of simple fluids and theory.--} We numerically simulate the
dynamics of periodic Lennard-Jones (LJ)~\cite{jones} and
Weeks-Chandler-Andersen (WCA)~\cite{weeks} fluids in three spatial dimensions.
The Hamiltonian of these $N$-particle systems is
$H(r_{ij},p_{k})=\sum_{k}^{3N}p_{k}^{2}/2m+\sum_{i<j}^{N}V(r_{ij})$. The
interparticle interactions are specified by the pairwise potential $V(r_{ij})$
between the particles $i$ and $j$ a distance $r_{ij}$ apart (inset of
Fig.~\ref{Fig6}(a)). In the WCA fluid, particles interact through short-range
repulsive forces and in the LJ fluid, particles also attract one another
through comparatively longer-range forces. All quantities are in reduced units.
(See Supplemental Material (SM)~\cite{supp} Secs. I and II.)

Along with each constant energy trajectory of these equilibrium fluids, we
simulate the corresponding tangent space dynamics, which limits the largest,
computationally tractable system size to around $2000$ particles, even
leveraging recent computational advances~\cite{cost}. Within the linearized
limit, the expansion or the compression factor of the perturbation along the
direction of the $i$th Lyapunov vector over time $t$ is $e^{\Gamma_{i}(t)}$.
The associated FTLE is $\lambda_{i}=\Gamma_{i}(t)/t$. For each trajectory, we
calculate the finite-time Lyapunov spectrum, $\{ \lambda_{i} \}$, and the
finite-time Kolmogorov-Sinai (KS) entropy $h_\textsc{\tiny KS}$ (i.e., the sum
of the positive $\{ \lambda_{i} \}$ using Pesin's theorem~\cite{pesin}) from
the set of Gram-Schmidt vectors~\cite{fuji,bene}. Algorithms for computing the
covariant Lyapunov vectors do not readily scale to the system sizes we
simulate here~\cite{gine}. 

Our interest is in the self-averaging of fluctuations of the FTLEs and the
finite-time KS entropy for simple fluids with and without attractive forces. We
quantify fluctuations in the FTLEs over fixed time intervals, $t$, with the
diffusion coefficients $\{ D(\lambda_{i})\}$~\cite{lopez,kupt,lopez1} and the
variance, $\chi_{i}^{2}(t)$, of $\{\Gamma_{i}(t)\}$
\begin{eqnarray}\label{2.5}
  tD(\lambda_{i}) = \chi_{i}^{2}(t) = \left<(\Gamma_{i}(t)- \left<\lambda_{i} \right>t )^{2} \right>. 
\end{eqnarray}
Averages $\left<\cdot\right>$ are over an ensemble of $10^{4}$ trajectories,
each with $t=0.1$ in reduced time units (SM~\cite{supp} Sec.~III).  $\left<
\lambda_{i}\right>$ is the long-time average of $\lambda_{i}$. With these
trajectory ensembles, we analyze the scaling of the diffusion coefficient for
the entire Lyapunov spectrum, $\{ D(\lambda_{i}) \}$, and the KS entropy rate,
$D(h_\textsc{\tiny KS})$, with the number of particles, $N$.

\textit{Self-averaging of first Lyapunov exponent.--} An important measure of
the degree of deterministic chaos in nonlinear dynamical systems is the
Lyapunov exponent associated with the maximally expanding tangent space
direction, $\lambda_{1}$. Over finite times, $\lambda_1$ concentrates around
the long-time average with increasing system size, $N$. Figure~\ref{Fig6}(a)
shows representative empirical distributions of $\lambda_{1}$ for the WCA fluid
over an ensemble of finite-time trajectories. We find similar results for the
LJ fluid.

\begin{figure}[b]
\includegraphics[width=\columnwidth,angle=0,clip]{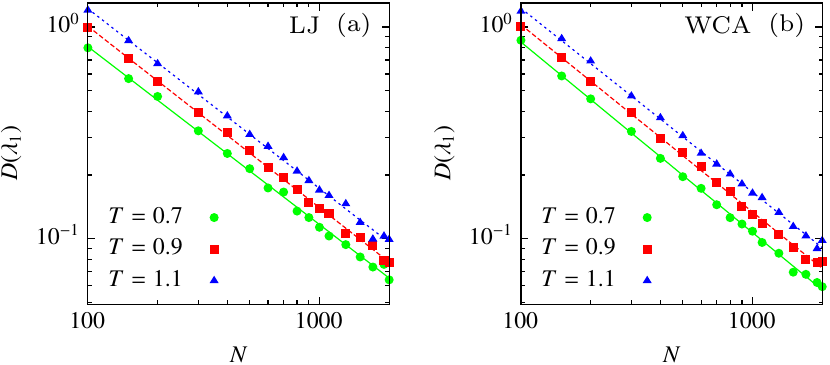}

\caption[]{\label{Fig1} Power law decay in the fluctuations of the finite-time
Lyapunov exponent, $\lambda_1$, as measured by the diffusion coefficient,
$D(\lambda_{1})$, with the number of particles, $N$. Here, $\lambda_1$ is the
exponent associated with the first Lyapunov vector. Data at three temperatures
are shown for the (a) Lennard-Jones and (b) Weeks-Chandler-Andersen fluids. The
points are simulation data and the lines are linear fits for systems with $N$
from $100$ to $2000$ particles.}

\end{figure}

To quantify the decay rate of Lyapunov exponent fluctuations, we scale the
number of molecules $N$ and the volume $V$ to ensure the thermodynamic limit of
the microcanonical ensemble: $N,V\to\infty$ keeping the number density
$\rho=N/V$ and energy density $e=E/V$ fixed. At constant $\rho$ and $e$, the
kinetic temperature is given by the equipartition theorem,
$T=2\left<E_\textrm{kin}\right>/3Nk_\textsc{\tiny B}$.

From scaling the system size of LJ and WCA fluids, we find that the diffusion
coefficient $D(\lambda_{1})$ scales as a power law, $\mathcal{O}
(N^{-\gamma})$, with wandering exponent $\gamma$ (Fig.~\ref{Fig1}).  The
magnitude of $\gamma$ depends on the nature of the intermolecular forces and,
in both cases, shows that fluctuations in $\lambda_{1}$ are weakly
self-averaging, $\gamma < 1$. For the WCA liquid, where molecules are purely
repulsive $\gamma \approx 0.9$. However, for the LJ liquid, where there are
also attractions over a few molecular diameters $\gamma \approx 0.85$. The
difference between these $\gamma$ values is outside our statistical errors and
shows that attractions act weakly to slow the decay of fluctuations.

Perturbative treatments of liquids in statistical mechanics build on the van
der Waals picture. They assume the structure of a dense, monatomic fluid
resembles that of a hard sphere fluid and, to a first approximation, the
attractive interactions have little effect on the liquid
structure~\cite{hansen}. Here, we see this picture through the fluctuations in
$\lambda_1$, which decay at different rates for the LJ and WCA fluids.
Repulsive forces dominate the decay of Lyapunov exponent fluctuations.
Attractive forces not only slow the divergence of trajectories~\cite{green1},
they also diminish the rate $\gamma$ at which the thermodynamic states of
liquids emerge through $\lambda_1$. These results are consistent with the idea
that subvolumes of liquids made up of repulsive particles will become
independent more quickly with increasing system size than those made of
particles that can also attract over a relatively longer range. They also
reinforce the intuition that the longer the range of intermolecular
interactions, the slower the rate at which the largest Lyapunov exponent will
self-average.

Although the weak self-averaging behavior of the largest Lyapunov exponent is
sensitive to the length scale of the intermolecular forces, it is not dependent
upon temperature. The wandering exponents for $\lambda_{1}$ of both fluids is
independent of temperature in the range $0.7-1.1$ (SM~\cite{supp} Sec.~IV).

\begin{figure}[t]
\includegraphics[width=0.7\columnwidth,angle=0,clip]{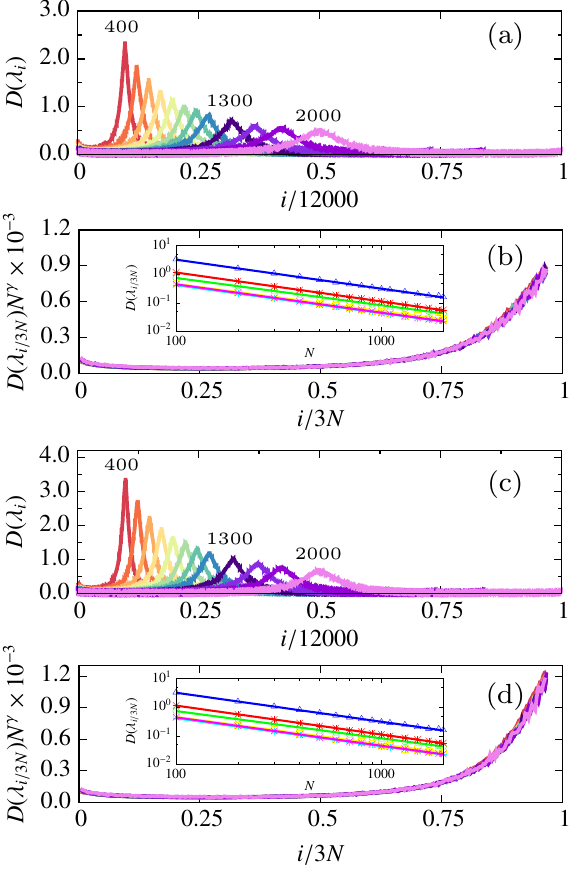}

\caption[]{\label{Fig2} The diffusion coefficient $D(\lambda_{i})$ as a
function of the scaled index $i/12000$ for (a) Lennard-Jones and (c)
Weeks-Chandler-Andersen fluids with $N=400-2000$ particles. The scale factor
normalizes the abscissa; $12000$ is the number of phase space dimensions for
the largest system. These data collapse upon rescaling the diffusion
coefficient $D(\lambda_{i/3N})N^{\gamma}$ as a function of the scaled index
$i/3N$ for both the (b) Lennard-Jones and (d) Weeks-Chandler-Andersen fluids.
There are $16$ systems in all, ranging from $100$ to $2000$ particles. Insets
in (b) and (d) show the scaling of $D(\lambda_{i/3N})$ with system size $N$ for
select scaled indices. The points are simulation data and the lines are linear
fits. In all cases, the kinetic temperature is $T=0.9$ and the number density
is $0.75$.}

\end{figure}

\textit{Self-averaging of the bulk Lyapunov exponents.--} The self-averaging of
the largest Lyapunov exponent is distinct from the bulk of the spectrum
(Fig.~\ref{Fig2}). And, for the fluids and conditions here, it measures the
relative importance of attractions and repulsions in the emergent dynamics of
simple liquids. Other observables, such as the kinetic and potential energy,
however, do not. Instead, the kinetic and potential energy fluctuations are
strongly self-averaging (SM~\cite{supp} Sec.~V) with $\gamma\approx 1$
independent of the intermolecular forces or temperature. Attractive forces also
do not measurably affect the self-averaging behavior of the bulk Lyapunov
exponents: all exponents in the spectrum beyond the first self-average
strongly, $\gamma \approx 1$. Previous work also found that the scaling of the
first Lyapunov exponent fluctuations with system size are markedly different
from the bulk of the spectrum in both conservative and dissipative,
one-dimensional dynamical systems~\cite{kupt,lopez1}.

Both the WCA and LJ fluids under $NVE$ conditions have symmetric Lyapunov
spectra~\cite{green1} as a result of the conjugate pairing
rule~\cite{morriss,posch} and the conservation of phase space volume by
Liouville's theorem. Figures~\ref{Fig2}(a) and~\ref{Fig2}(c) show this symmetry
in the diffusion coefficients of the Lyapunov spectra, $\{D(\lambda_{i})\}$,
for the LJ and WCA liquids. As a function of the spectral index $i$, the
diffusion coefficients are symmetric about half of the number of phase space
dimensions, $i=3N$~\cite{kupt}.

Figure~\ref{Fig2} shows representative results for the finite-size scaling of the
fluctuations in $\{ \lambda_{i} \}$. The insets in (b) and (d) show the
scaling of $D(\lambda_{i/3N})$ with $N$ for the LJ and the WCA liquids,
respectively, for several scaled indices, $i/3N$, at $T=0.9$. Our simulations
of $16$ systems span $100$ to $2000$ particles ($N=$ 100, 200, \ldots, 1100,
1300, \ldots, 1900, and 2000). The number of positive Lyapunov exponents and the
range of the index, $i$, grows as $3N$, so the scaled index, $i/3N$, is
intensive. Given the symmetry of $D(\lambda_{i})$, we focus on the positive
exponents $\{ \lambda_{i} \}$. We see that $\{ D(\lambda_{i/3N}) \}$ scales as
a power law, $N^{-\gamma}$ with $\gamma \approx 1$ for all exponents and both
the LJ and the WCA fluids (SM~\cite{supp} Sec.~VI). Furthermore, we find the
scaled diffusion coefficient $D(\lambda_{i/3N})N^{\gamma}$ data collapse onto a
single curve as a function of the scaled index.

While we model the fluids as three-dimensional Hamiltonian systems, we do not
detect the divergence of fluctuations of FTLEs in the thermodynamic limit seen
in one-dimensional Hamiltonian lattices, the FPU-$\beta$ and $\Phi^{4}$
models~\cite{lopez}.  The wandering exponent of the first Lyapunov exponent for
the FPU-$\beta$ model is $\gamma< -0.25$ and for the $\Phi^{4}$ model is
approximately $-1$. Paz\'o \textit{et al.} suggested that the divergence is
likely a result of long-range spatiotemporal correlations but could be a more
universal property of Hamiltonian dynamics.  However, our results suggest the
former is correct. Another consequence of these long-range correlations is that
the extensivity and additivity of the long-time average KS entropy does not
necessarily mean short-time fluctuations of the bulk Lyapunov exponents will be
strongly self-averaging as we see here. Although the length scale of LJ
interactions (of the order of a few molecular diameters) is longer than the
purely repulsive WCA interactions, it is still sufficiently short range for the
Lyapunov exponents to self-average.

The rate of self-averaging of Lyapunov exponent fluctuations depends on the
number of spatial dimensions. Weak self-averaging of $\lambda_{1}$ for one- and
two-dimensional systems \textit{with weak correlations} is characterized by a
$\gamma$ of $1/2$ and $0.839$, respectively~\cite{kupt,lopez1}. The wandering
exponents for the three-dimensional LJ and WCA systems are higher, $0.85$ and
$0.9$. Because the KPZ critical exponents are known in lower dimensions,
dynamical systems can be assigned to the KPZ universality class. But, this is
not the case in three dimensions, so an assignment for the LJ and WCA fluids is
not currently possible. Still, the agreement between the wandering exponents in
fluids with and without attractive forces is evidence that the bulk of the
Lyapunov vectors do belong to the same (albeit unknown) universality class,
despite the longer range of the LJ interaction potential compared to WCA\@.

Strong self-averaging of the bulk Lyapunov exponents is so far unique to these
simple fluids. In one-dimensional systems with short-range correlations, for
example, the bulk Lyapunov spectrum self-averages weakly with $\gamma \approx
0.85$~\cite{kupt,lopez1}. These lower dimensional systems also have an
extensive and additive KS entropy, so this comparison also suggests that the
strong self-averaging of the bulk exponents in simple fluids does not
necessarily lead to these properties of the long-time KS entropy~\cite{green1}.
While it is well known from the van der Waals picture that repulsions dictate
liquid structure, our results show that this picture is also apparent in the
fluctuations of the Lyapunov exponents. Attractions have a measurable effect on
only the most unstable Lyapunov direction.

\begin{figure}
\includegraphics[width=\columnwidth,angle=0,clip]{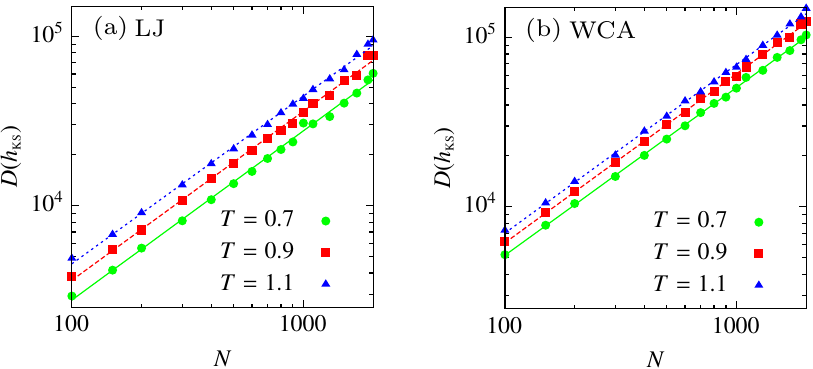}

\caption[]{\label{Fig4} Fluctuations in the finite-time Kolmogorov-Sinai
entropy rate, as measured by the associated diffusion coefficient,
$D(h_{\textsc{\tiny KS}})$, diverge with $N$. Data at three temperatures are
shown for the (a) Lennard-Jones and (b) Weeks-Chandler-Andersen fluids. The
points are simulation data and the lines are linear fits for $N$ from $100$ to
$2000$ particles.}

\end{figure}

\textit{Self-averaging of the Kolmogorov-Sinai entropy rate and
fluctuation-dissipation relations.--} Another important quantity characterizing
dynamical systems is the finite-time KS entropy: the sum of the positive FTLEs,
$h_\textsc{\tiny KS}(t)=\sum_{i}^{+} \lambda_{i}(t)$. The fluctuations in the
KS entropy, though important in fluctuation theorems, are less explored
compared to those of the FTLEs. Again, we measure the finite-time KS entropy
fluctuations with the diffusion coefficient. The scaling of $D(h_\textsc{\tiny
KS})$ with system size is a power law $N^{+\gamma}$ with $\gamma \approx 1$, as
shown in Figs.~\ref{Fig4}(a) and (b), for both the LJ and the WCA fluids. The
divergence of fluctuations in the KS entropy supports recent evidence of its
system-size extensivity; the converging Lyapunov exponent fluctuations also
confirm their system-size intensivity~\cite{green1}. Because the KS entropy for
these fluids is system-size extensive, fluctuations in the KS entropy density,
$h_\textsc{\tiny KS}/N$, are self-averaging and decay as $N^{-1}$ to the
thermodynamic limit. 

As for the Lyapunov exponents, temperature does not affect the scaling
(SM~\cite{supp} Sec.~IV). However, the magnitude of the diffusion coefficient
is temperature dependent. To determine the temperature dependence, we ran
additional simulations for the LJ and the WCA fluids between $T=0.4$ and $2.0$.
From these simulations, for all the Lyapunov exponents, the first and the bulk,
and the Kolmogorov-Sinai entropy for both Lennard-Jones and
Weeks-Chandler-Andersen fluids, $D$ varies as $k_BT$ over the temperatures and
numbers of particles we consider (SM~\cite{supp} Sec.~IV).  Together with the
self-averaging of these dynamical observables, the linear dependence of their
diffusion coefficients on temperature suggest an empirical
fluctuation-dissipation relation for chaoticity, $D\propto k_BT N^{\pm\gamma}$,
akin to the Einstein-Smoluchowski relation in real space, $D = \mu k_BT$.
These results, however, connect the rate at which trajectories chaotically
diffuse through phase space and the equilibrium temperature.

In summary, the dominance of repulsive forces over attractive forces in the
dynamics of fluids is manifest in the fluctuations of the Lyapunov spectrum. We
showed the first Lyapunov exponent self-averages, but only weakly, for
three-dimensional Hamiltonian models of simple fluids. Weak attractions merely
act to slow the rate of decay of fluctuations scaling to the thermodynamic
limit. The convergence of the fluctuations associated with the first Lyapunov
exponent for the three-dimensional Hamiltonian systems here is distinct from
the divergent behavior of the one-dimensional Hamiltonian models studied to
date. We attribute this difference to the short-range interparticle forces in
the WCA and LJ fluids. These forces affect the wandering exponent of
$\lambda_{1}$ but not the strongly self-averaging nature of the bulk exponents.
Consequently, the fluctuations in the KS entropy diverge as one expects for an
extensive thermodynamic quantity.

From the wandering exponents, the bulk Lyapunov vectors of Lennard-Jones and
Weeks-Chandler-Andersen fluids belong to the same, as yet unknown, universality
class. What is clear is that the classes of the first Lyapunov vectors are
distinct. It remains to be seen whether the spatial dimension, the length scale
of interparticle interactions, or both influences the rate at which
fluctuations decay in the bulk of the spectrum. Still, it is tempting to
speculate that increasing the length scale of attractive forces would further
slow the convergence and, eventually, even cause fluctuations to diverge.
Regardless of intermolecular forces, for simple equilibrium liquids, we find
evidence that the spontaneous fluctuations of these dynamical observables obey
fluctuation-dissipation like relationships. Altogether, our results give an
alternate view of the van der Waals picture of liquids, a view that opens up
the possibility of examining emergent dynamical signatures at the liquid-gas
critical point, in viscous liquids, or in self-organizing
systems~\cite{greencgs}, where we anticipate long-range correlations to cause
nontrivial self-averaging behavior.

Acknowledgment is made to the donors of The American Chemical Society Petroleum
Research Fund for support of this research. We acknowledge the use of the
supercomputing facilities managed by the Research Computing Group at the
University of Massachusetts Boston. We thank M.~Alaghemandi and S.~B.~Nicholson
for useful discussions.

\end{document}